\documentclass[useAMS,usenatbib]{mn2e}
\usepackage{graphicx}
\usepackage{rotfloat}

\newcommand{\ms}{$\,$M$_\mathrm{\odot}$}

\newcommand{\be}{\begin{equation}}
\newcommand{\ee}{\end{equation}}
\newcommand{\stars}{{\sc stars}}
\newcommand{\el}[2]{\ensuremath{^{#1}\mathrm{#2}}}

\title[Extra mixing in metal-poor giants]{The depletion of carbon by extra mixing in metal-poor giants}
\author[R.~J. Stancliffe, R.~P. Church, G.~C. Angelou \& J.~C. Lattanzio]{Richard J. Stancliffe\thanks{E-mail:
Richard.Stancliffe@sci.monash.edu.au}, Ross P. Church, George C. Angelou and John C. Lattanzio\\
Centre for Stellar and Planetary Astrophysics, Monash University, PO Box 28M, Victoria 3800, Australia \\
}
\begin{document}
\bibliographystyle{mn2e}

\date{Accepted 0000 December 00. Received 0000 December 00; in original form 0000 October 00}

\pagerange{\pageref{firstpage}--\pageref{lastpage}} \pubyear{0000}

\maketitle

\label{firstpage}

\begin{abstract}
There is an apparent dichotomy between the metal-poor ([Fe/H]$\ \leq -2$) yet carbon-normal giants and their carbon-rich counterparts. The former undergo significant depletion of carbon on the red giant branch after they have undergone first dredge-up, whereas the latter do not appear to experience significant depletion. We investigate this in the context that the extra mixing occurs via the thermohaline instability that arises due to the burning of \el{3}{He}. We present the evolution of [C/Fe], [N/Fe] and \el{12}{C}/\el{13}{C} for three models: a carbon-normal metal-poor star, and two stars that have accreted material from a 1.5\ms\ AGB companion, one having received 0.01\ms\ of material and the other having received 0.1\ms. We find the behaviour of the carbon-normal metal-poor stars is well reproduced by this mechanism. In addition, our models also show that the efficiency of carbon-depletion is significantly reduced in carbon-rich stars. This extra-mixing mechanism is able to reproduce the observed properties of both carbon-normal and carbon-rich stars. 
\end{abstract}


\section{Introduction}

The pursuit of the most metal-poor stars in the Universe is one of the most active fields of research in modern stellar astronomy. Several large-scale surveys, such as the `First Stars' program of Cayrel and collaborators \citep[e.g.][]{2004A&A...416.1117C}, the Hamburg/ESO survey of Christlieb and collaborators \citep[e.g.][]{2003RvMA...16..191C} and the HK survey of Beers and collaborators \citep[e.g.][]{1992AJ....103.1987B} have yielded a wealth of abundance determinations. This accumulation of data provides us with statistically meaningful populations that can be compared to theoretical predictions. One result of such survey work is that two populations of stars can be defined based on their carbon abundances. There is a distinct separation between those stars that are carbon-rich and those that are carbon-normal. The dividing line is drawn at [C/Fe]\footnote{[A/B] = $\log (N_\mathrm{A}/N_\mathrm{B}) - \log (N_\mathrm{A}/N_\mathrm{B})_\odot$}$>$+1.0 \citep{2005ARA&A..43..531B} and the carbon-rich population accounts for a substantial fraction of all the metal-poor stars \citep[e.g.][]{2006ApJ...652L..37L}.

\begin{figure}
\includegraphics[width=\columnwidth]{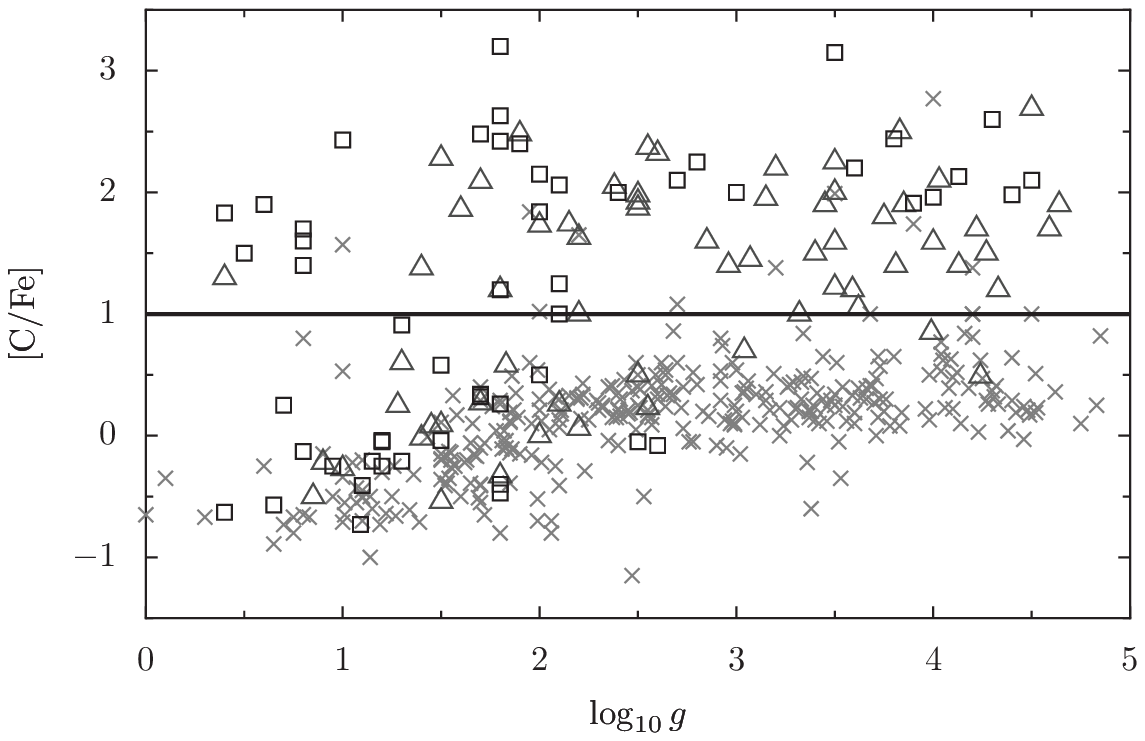}
\caption{[C/Fe] as a function of $\log_{10} g$ for metal-poor stars extracted from the SAGA database \citep{2008arXiv0806.3697S}. Light grey crosses represent stars that only have [C/Fe] measured, while dark open triangles represent stars that have measurements of [C/Fe] and [N/Fe]. The black open squares are those stars for which we have data for [C/Fe], [N/Fe] and \el{12}{C}/\el{13}{C}. The solid line at [C/Fe]~=~1 denotes the dividing line between carbon-normal and carbon-rich metal-poor stars.}
\label{fig:gCFe}
\end{figure}

There is an interesting dichotomy between the behaviour of carbon-normal and carbon-enhanced metal-poor (CEMP) stars. This is shown in Figure~\ref{fig:gCFe} which shows measurements of [C/Fe] as a function of surface gravity, $\log_{10} g$. Stars have higher surface gravities when they are more compact. On the main sequence $\log_{10} g\approx5$ and as the star expands, $\log_{10} g$ drops. By the time $\log_{10} g\approx3$, the star has reached the giant branch and its convective envelope deepens, bringing CN-cycled material to the surface. This is first dredge-up. The star continues to expand and by the tip of the red giant branch, $\log_{10} g\approx0.5$. The carbon-normal stars, once they reach a surface gravity of around $\log_{10} g\approx2$ (i.e. stars that have passed through first dredge-up and are on the upper part of the red giant branch), show a marked depletion in their surface carbon abundance. [C/Fe] drops by nearly one dex between $\log_{10} g\approx2-0.5$. On the basis of this drop in the [C/Fe] abundance, \citet{2007ApJ...655..492A} suggested that the definition of carbon-enhancement should be revised to take account of this depletion. They suggest that the appropriate criteria should be:
\begin{enumerate}
\item{[C/Fe]$\ \geq +0.7$, for stars with $\log\ (L/\mathrm{L_\odot})\leq2.3$}
\item{[C/Fe]$\ \geq + 3.0 - \log\ (L/\mathrm{L_\odot})$, for stars with $\log\ (L/\mathrm{L_\odot})>2.3$}
\end{enumerate}

While there is clear evidence for the depletion of carbon in the C-normal population, the carbon-rich metal-poor stars show no sign of carbon depletion when $\log_{10} g<2$. This was noted by \citet{2008ApJ...679.1541D}. In fact, there is a noticeable dearth of stars in the triangular regions whose apexes are at ($\log_{10} g$, [C/Fe]) = (2,1), (0.5,1), (0.5,-0.5). This is unlikely to be a selection effect as observations are sensitive to this region of parameter space \citep{2007ApJ...667.1185L}. It therefore seems that carbon-enhanced metal-poor stars do not suffer the same degree of carbon-depletion on the upper part of the red giant branch as their carbon-normal cousins do.

The existence of extra mixing in red giant branch stars is observationally well established.  \citet{1975ApJ...200..675D}, \citet{1976ApJ...210..694T} and more recently \citet{2001AJ....122.2554K} used \el{12}{C}/\el{13}{C} ratios as an indicator of the efficiency of mixing associated with first dredge up. The discrepancy between canonical models and observations were the first indications that stars might undergo some form of extra mixing. Further work by \citet{1998A&A...332..204C} and \citet{2008arXiv0810.1701S} showed that there are unexplained increases in N whilst \el{7}{Li}, C and \el{12}{C}/\el{13}{C} all decrease. Systematic studies of various stellar environments -- including field stars, globular clusters, open clusters and the Large Magellanic Cloud -- have shown evidence of extra mixing operating universally across a range of masses and metallicities \citep[see][and references therein]{2008arXiv0810.1701S}. The efficiency of the mixing has also been well studied with \citet{2001AJ....122.2554K} determining mixing is more efficient at lower metallicities.  \cite{2008arXiv0809.4470M} show progressive C depletion with increasing luminosity along the giant branch suggesting that the mixing occurs throughout a star's ascent of the giant branch.  

The mechanism for this additional mixing has proved somewhat elusive. Extra mixing caused by rotation was proposed by \citet{1973AcA....23..191P} and again by \citet{1978ApJ...221..893C} to explain low C/N ratios in young stars and weak G band stars respectively. \citet{1979ApJ...229..624S} were the first to investigate the role of meridional circulation on the RGB and this has been an avenue of pursuit for many authors since. \citet{2000MNRAS.316..395D}, applying the formalism of \citet{1998A&A...334.1000M} in a post-processing approach, showed that rotational mixing could produce abundance variations on the RGB. However, their models also produced anticorrelations in O-Na and Mg-Al. It is now thought that these features are not caused by extra mixing on the giant branch, as they have been observed in turn-off stars in globular clusters. \citet{2003ApJ...593..509D} subsequently improved on this work, showing that the abundance anomalies could be reproduced if the diffusion coefficient from their rotating model was multiplied by a factor of 7. However, the recent models by \citet{2006A&A...453..261P} cast serious doubt on the assumption that rotation is the mechanism responsible for extra mixing on the RGB. Using models in which the transport of angular momentum by meridional circulation and shear turbulence, in addition to the associated chemical mixing these processes cause, was self-consistently treated (i.e. the feedback between rotational processes and stellar structure was taken into account), these authors showed that rotational mixing was unable to account for the observed abundance changes \citep[see figure 15 in][]{2006A&A...453..261P}. 

Recently, an alternative mechanism has received much attention as the possible cause of mixing on the RGB. The direct numerical simulations of \citet{2006Sci...314.1580E} showed that the reaction \el{3}{He}(\el{3}{He},2p)\el{4}{He} \citep[this reaction and its effect upon thermohaline mixing in pre-main sequence giants was previously studied by][]{1972ApJ...172..165U} could lead to a lowering of the mean molecular weight above the H-burning shell and that this would lead to mixing. In follow-up work, these authors modelled this process via a diffusion coefficient which gave diffusive velocities in the range expected from simple physical arguments. They then went on to show how this could affect the isotopic ratios of various elements \citep{2008ApJ...677..581E}. As a star evolves up the giant branch, its convective envelope deepens and material that has undergone CN-cycling is brought to the surface. This is the process known as first dredge-up. First dredge-up homogenises the stellar envelope, resulting in a uniform mean molecular weight $\mu$. As the hydrogen burning shell reaches this homogenous region \el{3}{He} starts to burn, lowering the mean molecular weight in this region. The resultant $\mu$-profile is unstable to thermohaline mixing\footnote{We use the terms thermohaline mixing and $\delta \mu$ mixing interchangeably throughout this paper.} and the material begins to be circulated between the burning region and the convective envelope. This connection of the burning region to the envelope (and hence the stellar surface) causes the surface abundances to change, with carbon becoming depleted and nitrogen enhanced. One particularly appealing aspect of this mechanism is that it is a direct consequence of the physics of nuclear burning and does not require a particular rotation rate or rotational profile (or even magnetic field strength, if such physics were to be included) -- hence the reason that Eggleton and collaborators dubbed the process ``compulsory'' \citep{2008ApJ...677..581E}.

The effect of this `$\delta\mu$ mixing' has been investigated for a range of metallicities \citep{2008ApJ...677..581E,2007A&A...467L..15C}, with the general conclusion that mixing is more efficient at lower metallicity, in line with observations. In this paper, we apply this mechanism to both the carbon-rich and the carbon-normal metal-poor populations to see what effects it may have.

\section{The stellar evolution code}
Calculations in this work have been carried out using a modified version of the \stars\ stellar evolution code originally developed by \citet{1971MNRAS.151..351E} and updated by many authors \citep[e.g.][]{1995MNRAS.274..964P}. The code solves the equations of stellar structure and chemical evolution in a fully simultaneous manner, iterating on all variables at the same time in order to converge a model \citep[see][for a detailed discussion]{2006MNRAS.370.1817S}. The version used here includes the nucleosynthesis routines of \citet{2005MNRAS.360..375S} and \citet{stancliffe05}, which follow the nucleosynthesis of 40 isotopes from D to \el{32}{S} and important iron group elements. The code uses the opacity routines of \citet{2004MNRAS.348..201E}, which employ interpolation in the OPAL tables \citep{1996ApJ...464..943I} and which account for the variation in opacity as the C and O content of the material varies. In addition, molecular opacities are accounted for as described by \citet{2008MNRAS.389.1828S}. We note that metal-poor stars are typically $\alpha$-enhanced and so the use of opacity tables computed with such a composition would be appropriate. However, tables with both $\alpha$-enhancement and variable C and O content do not currently exist.

We have evolved three models, all with a metallicity of $Z=10^{-4}$ ([Fe/H]$\ \approx-2.3$): one is a standard carbon-normal, metal-poor model with solar-scaled abundances \citep{1989GeCoA..53..197A} while the other two are post-accretion models taken from \citet{2008MNRAS.389.1828S} and \citet{2009MNRAS.394.1051S}. The former has been evolved from the pre-main sequence and has a mass of 0.8\ms. The latter two have post-accretion masses of 0.8\ms, after having accreted 0.1 and 0.01\ms\ of material (at a rate of $10^{-6}\mathrm{M_\odot yr^{-1}}$ which is roughly equivalent to the rate of accretion expected in a wind mass transfer scenario) from a 1.5\ms\ AGB companion. These two models were evolved using thermohaline mixing and as such the accreted material has been mixed to equilibrium, with accreted material reaching a depth of 0.6 and 0.33\ms\ from the surface respectively \citep[see][for further details of these  models]{2008MNRAS.389.1828S}. The mixing reaches equilibrium before the end of the main sequence. In each case, the models have been remeshed to 999 meshpoints immediately after first dredge-up. Thermohaline mixing has been included via the prescription of \citet*{1980A&A....91..175K}, with the diffusion coefficient multiplied by a factor of 1000, following \citet{2007A&A...467L..15C}. Note that this is not consistent with the models of \citet{2008MNRAS.389.1828S}, who use just the \citet{1980A&A....91..175K} prescription for the mixing of accreted material. This is unlikely to effect the extent to which accreted material is mixed during the main sequence as the timescale for this processes is already significantly faster than the nuclear timescale that governs the star's evolution at this point.

\section{Results}

Before embarking on a description of the results of our simulations it is necessary to discuss the sample of observations against which we are going to compare our models. We have obtained our sample from the Stellar Abundances for Galactic Archaeology (SAGA) database \citep{2008arXiv0806.3697S}, selecting those stars which have [Fe/H]$\leq-2$ and for which [C/Fe] has been measured. This gives us a sample of 621 stars. Not all these stars have measurements of [N/Fe] and \el{12}{C}/\el{13}{C}, so we subdivide our sample into 3 sets: those stars for which we only have a [C/Fe] determination (444 stars), those stars for which we have both [C/Fe] and [N/Fe] but no \el{12}{C}/\el{13}{C} (122 stars) and finally those stars for which we have a determination for [C/Fe], [N/Fe] and \el{12}{C}/\el{13}{C} (55 stars). One note of caution should be added: the stars we have selected come from a range of different surveys and so do not represent a homogenous sample. This is forced on us of necessity -- we are collecting together data that was originally obtained for other purposes. There does not exist (to the knowledge of the authors) a single, homogenous set of data that measures all the abundances we require and yet covers both the carbon-normal and the carbon-enhanced metal-poor stars. 

\begin{figure}
\includegraphics[width=\columnwidth]{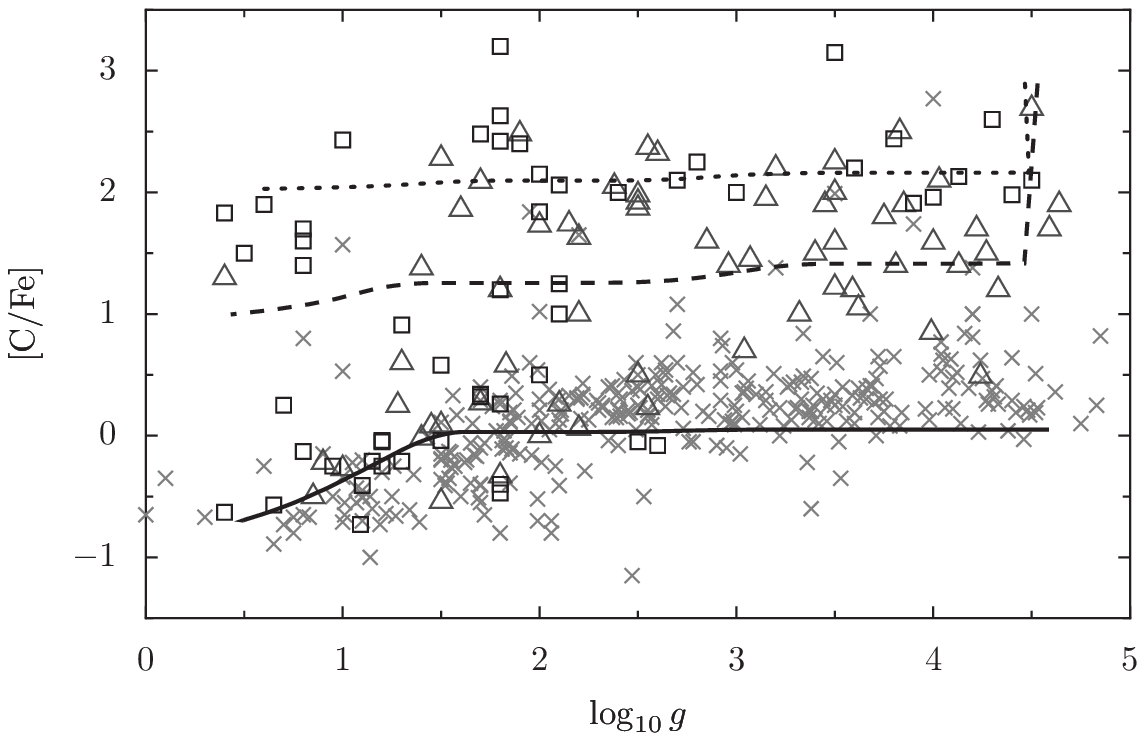}
\caption{The evolution of the [C/Fe] abundance as a function of $\log_{10} g$ for three models. The solid line is for a standard metal-poor, carbon-normal model. The other two models have accreted material from a 1.5\ms\ companion and this material has been allowed to mix via the thermohaline instability. The dotted line is for the case that 0.01\ms\ of material has been accreted, while the dashed line is for the case that 0.1\ms\ has been accreted. The initial drop in [C/Fe] at around $\log_{10} g\approx 4.5$ is caused by thermohaline mixing acting on the accreted material. Crosses represent objects for which we only have [C/Fe], while open triangles represent objects for which we have [C/Fe] and [N/Fe]. Objects for which [C/Fe], [N/Fe] and \el{12}{C}/\el{13}{C} are known are denoted by open squares.}
\label{fig:CFeModels}
\end{figure}

The results of the model runs are displayed in Figure~\ref{fig:CFeModels}. The standard carbon-normal model agrees very well with the data for stars with [C/Fe]$\ <\ 1$. Its [C/Fe] abundance falls by about 0.73 dex due to the effect of $\delta\mu$ mixing and this process sets in at $\log_{10} g\approx1.5$, just as is observed in the data. The model which has accreted 0.01\ms\ shows a much shallower drop in its carbon abundance with a change of about 0.4 dex. The model which has accreted 0.1\ms\ shows almost no change (less than 0.08 dex) in [C/Fe] as it evolves to lower surface gravities. 

\begin{figure}
\includegraphics[width=\columnwidth]{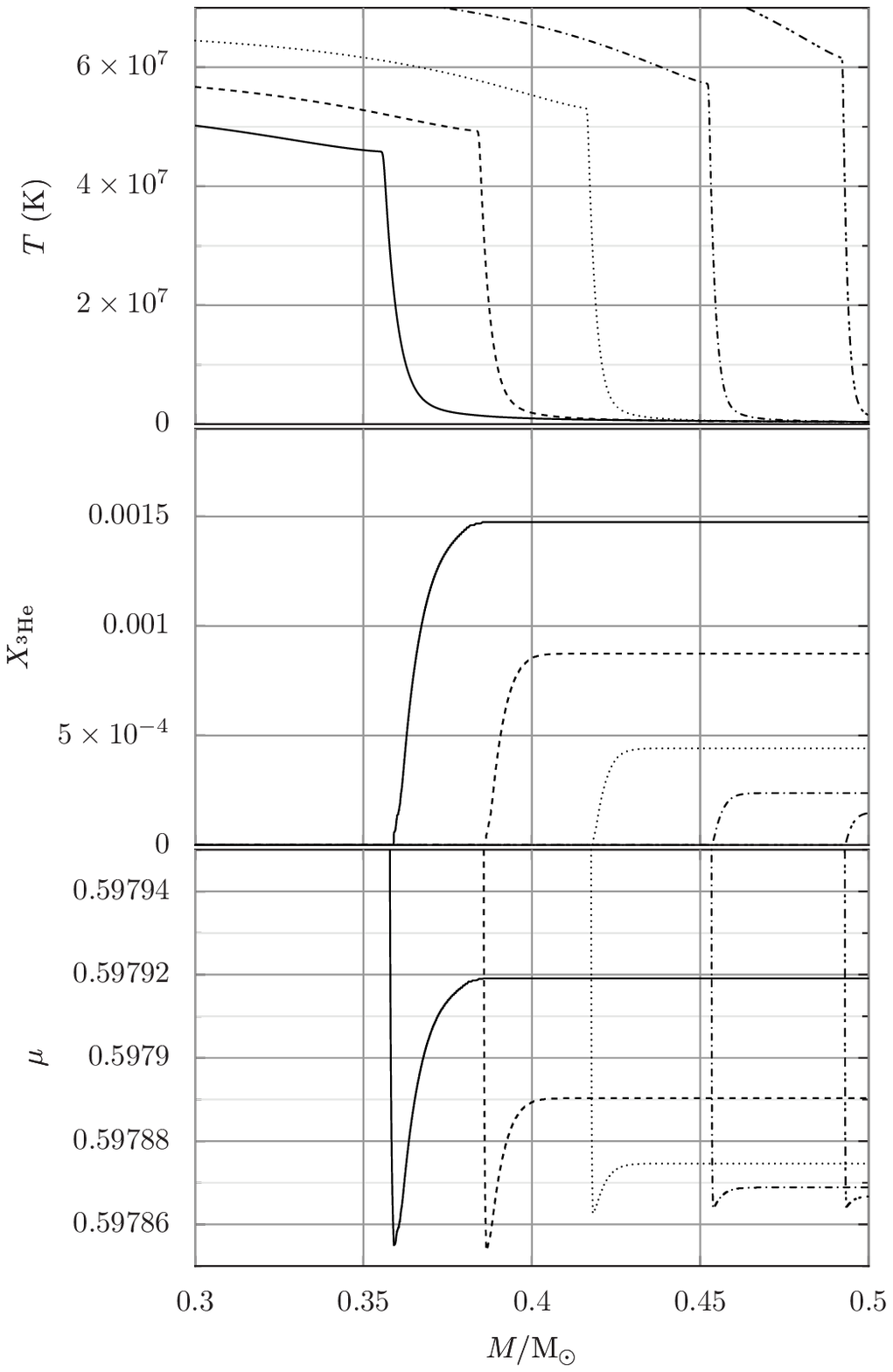}
\caption{Profiles as a function of mass for the standard carbon-normal model. The panels display the temperature profile (top panel), \el{3}{He} abundance (middle panel) and $\mu$ (bottom panel) as the model evolves. The solid line is the earliest model, representing the point at which $\delta\mu$ mixing begins to affect the surface abundances.}
\label{fig:Profiles_noacc}
\end{figure}

\begin{figure}
\includegraphics[width=\columnwidth]{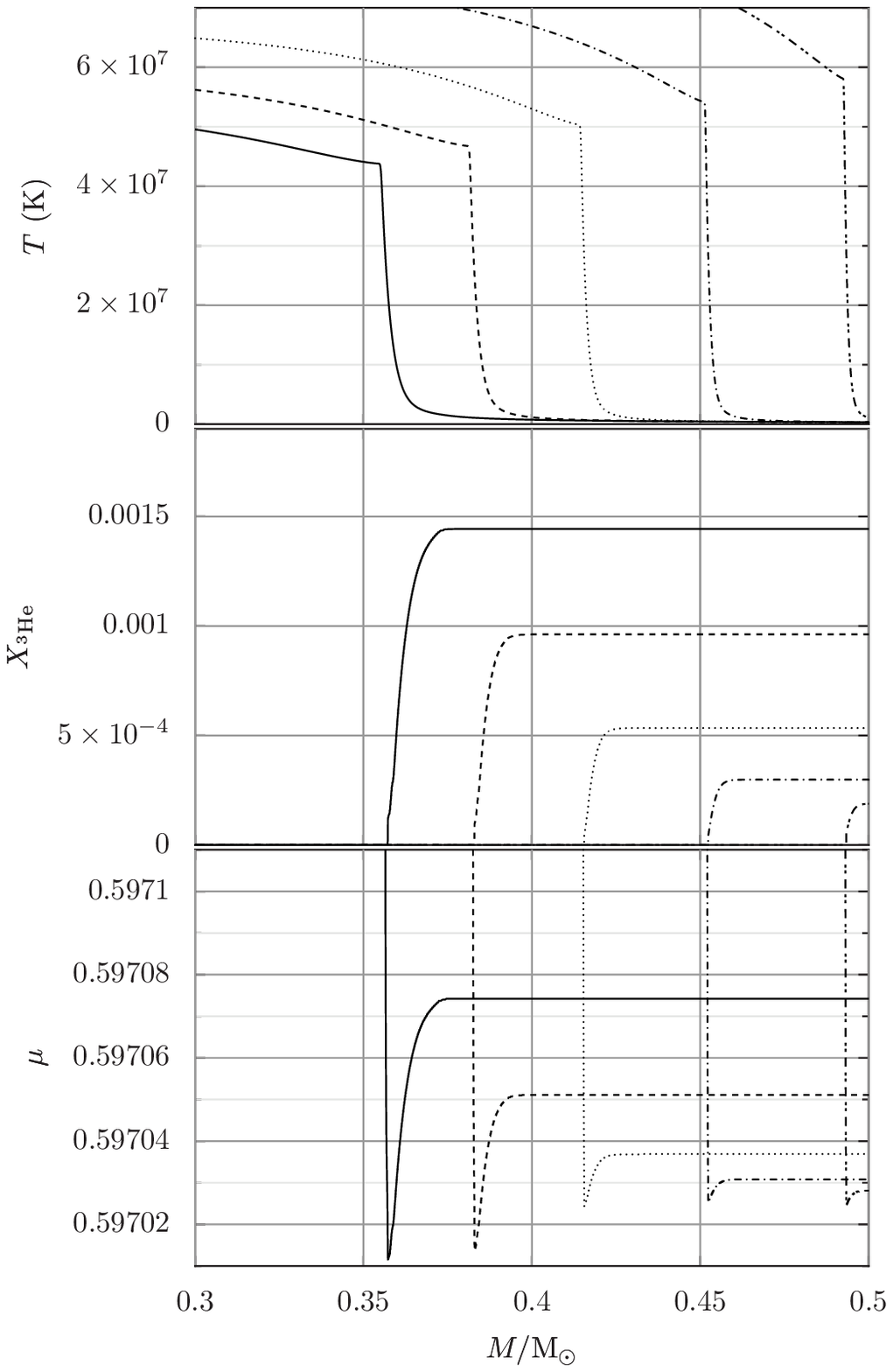}
\caption{Profiles as a function of mass for the carbon-rich model in which 0.01\ms\ of material was accreted. The panels display the temperature profile (top panel), \el{3}{He} abundance (middle panel) and $\mu$ (bottom panel) as the model evolves. The solid line is the earliest model, representing the point at which $\delta\mu$ mixing begins to affect the surface abundances.}
\label{fig:Profiles_0.01}
\end{figure}

\begin{figure}
\includegraphics[width=\columnwidth]{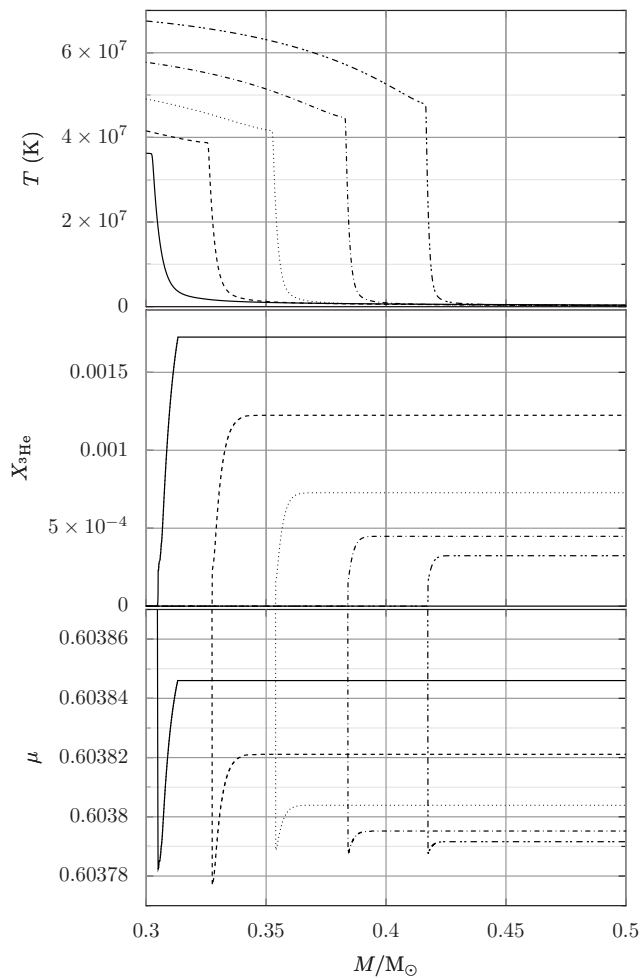}
\caption{Profiles as a function of mass for the carbon-rich model in which 0.1\ms\ of material was accreted. The panels display the temperature profile (top panel), \el{3}{He} abundance (middle panel) and $\mu$ (bottom panel) as the model evolves. The solid line is the earliest model, representing the point at which $\delta\mu$ mixing begins to affect the surface abundances.}
\label{fig:Profiles_0.1}
\end{figure}

The variation in behaviour can be explained by looking at the evolution of the interiors of these models, particularly at the temperature, $\mu$ and \el{3}{He} abundance profiles. These are displayed in Figures~\ref{fig:Profiles_noacc}, \ref{fig:Profiles_0.01} and \ref{fig:Profiles_0.1}. In each case, a similar reservoir of \el{3}{He} is available and the dip in $\mu$ caused by \el{3}{He} burning is roughly the same. Comparing the carbon-normal model (Figure~\ref{fig:Profiles_noacc}) with the 0.01\ms\ accretion case (Figure~\ref{fig:Profiles_0.01}), we see that the two models have almost identical temperature profiles with the $\mu$ dip occurring at a mass co-ordinate of about 0.36\ms. 
However, the 0.01\ms\ accretion case evolves to the tip of the giant branch faster than the carbon-normal model. It behaves like a star of higher metallicity and has a smaller core mass at the point of helium ignition. Evolution from the onset of $\delta\mu$ mixing to the tip of the giant branch takes $1.75\times10^7$\,yr in the carbon-normal model, while in the case of the 0.01\ms\ accretion model, this time is just $1.4\times10^{7}$\,yr. Carbon depletion is more efficient at later times when the temperature at the base of the mixing region is higher, so although the time spent undergoing mixing differs by 20 per cent the difference in the level of carbon depletion is much greater, by a factor of about 2.

One might assume that this argument would extend to the model in which 0.1\ms\ of material has been accreted. It has an an even larger carbon abundance in its envelope and so one might expect the core to grow faster and the star to spend even less time undergoing $\delta\mu$ mixing. However, this is not the case: the evolution for the onset of $\delta\mu$ mixing to the tip of the giant branch takes $2.4\times10^7$\,yr. We must look more closely at the structure of this model if we are to understand why this model suffers less carbon depletion. Firstly, we note that the core mass at which mixing commences in this model is much smaller than in the previous two cases. In the 0.1\ms\ accretion model, mixing begins at a core mass of about 0.3\ms, compared with 0.35\ms\ in the other two cases. This then affects the temperature structure: it is significantly cooler in the 0.1\ms\ case (see Figure~\ref{fig:Profiles_0.1}). The reduced temperature means that less CN cycling occurs in the layers of the star that undergo mixing and hence there is less of a drop in the carbon abundance.

\begin{figure}
\includegraphics[width=\columnwidth]{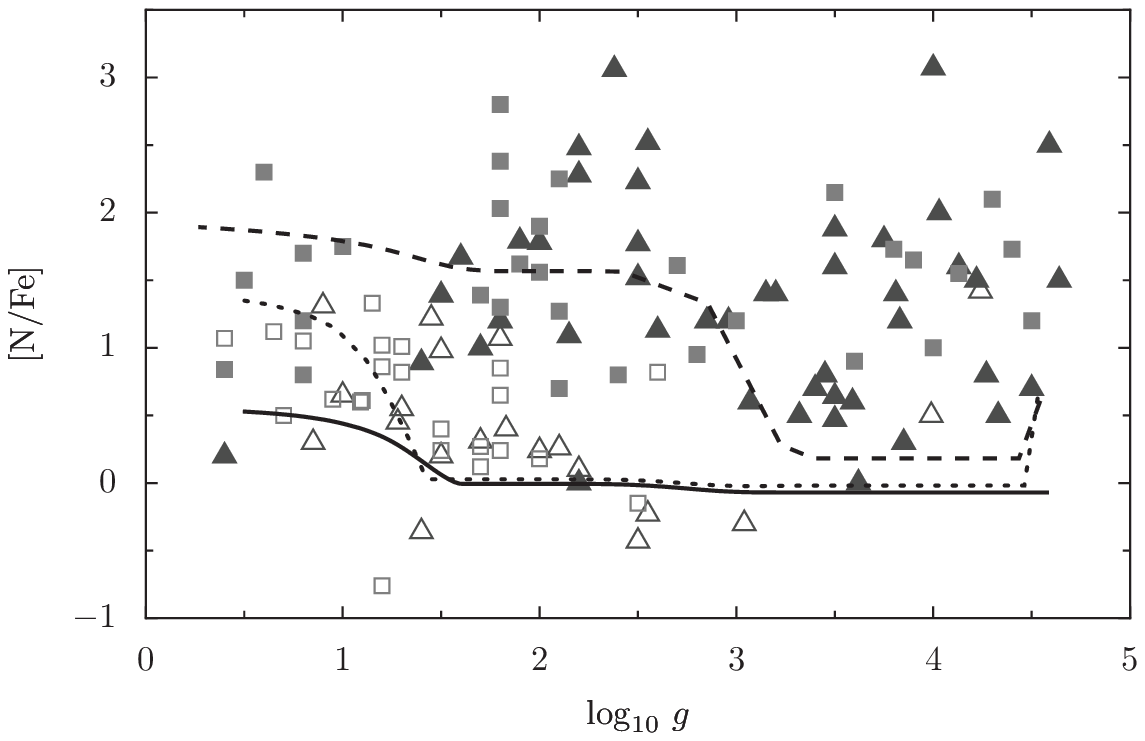}
\caption{The evolution of the [N/Fe] abundance as a function of $\log_{10} g$ for three models. The solid line is for a standard metal-poor, carbon normal model. The other two models have accreted material from a 1.5\ms\ companion and this material has been allowed to mix via the thermohaline instability. The dotted line is for the case that 0.01\ms\ of material has been accreted, while the dashed line is for the case that 0.1\ms\ has been accreted. Note that the drop in abundance at $\log_{10} g\approx 4.5$ occurs as the accreted material is mixed into the star via thermohaline mixing. Triangles represent those objects for which we only have [C/Fe] and [N/Fe], while squares represent those objects for which we also have the \el{12}{C}/\el{13}{C} ratio. Open symbols are for C-normal objects and filled symbols denote C-rich objects.}
\label{fig:NFe}
\end{figure}

Figure~\ref{fig:NFe} shows the evolution of [N/Fe] for our models alongside the [N/Fe] measurements for our sample of stars extracted from the SAGA database \citep{2008arXiv0806.3697S}. We note that the 0.1\ms\ model, which underwent extensive mixing on the main sequence, shows an extremely large increase in its nitrogen abundance at first dredge-up ($\log_{10} g\approx3$). This is because the material that was accreted has been mixed to a depth of around 0.6\ms\ from the surface. At this depth the temperature is high enough for CN-cycling to occur, converting the accreted C into N during the main sequence  \citep{2007A&A...464L..57S}. This material is then brought to the surface at first dredge-up. Each of the model sequences then shows a rise in the nitrogen abundance when the $\delta\mu$ mixing occurs. The final nitrogen abundance is proportional to the quantity of carbon available and hence the carbon-normal model ends up being the most nitrogen poor. We note that the models do not cover the full spread in nitrogen abundances, particularly at high surface gravities (i.e. on the main-sequence and at turn-off). However, the production of nitrogen from carbon does not exceed the observed values in any of our models.

\begin{figure}
\includegraphics[width=\columnwidth]{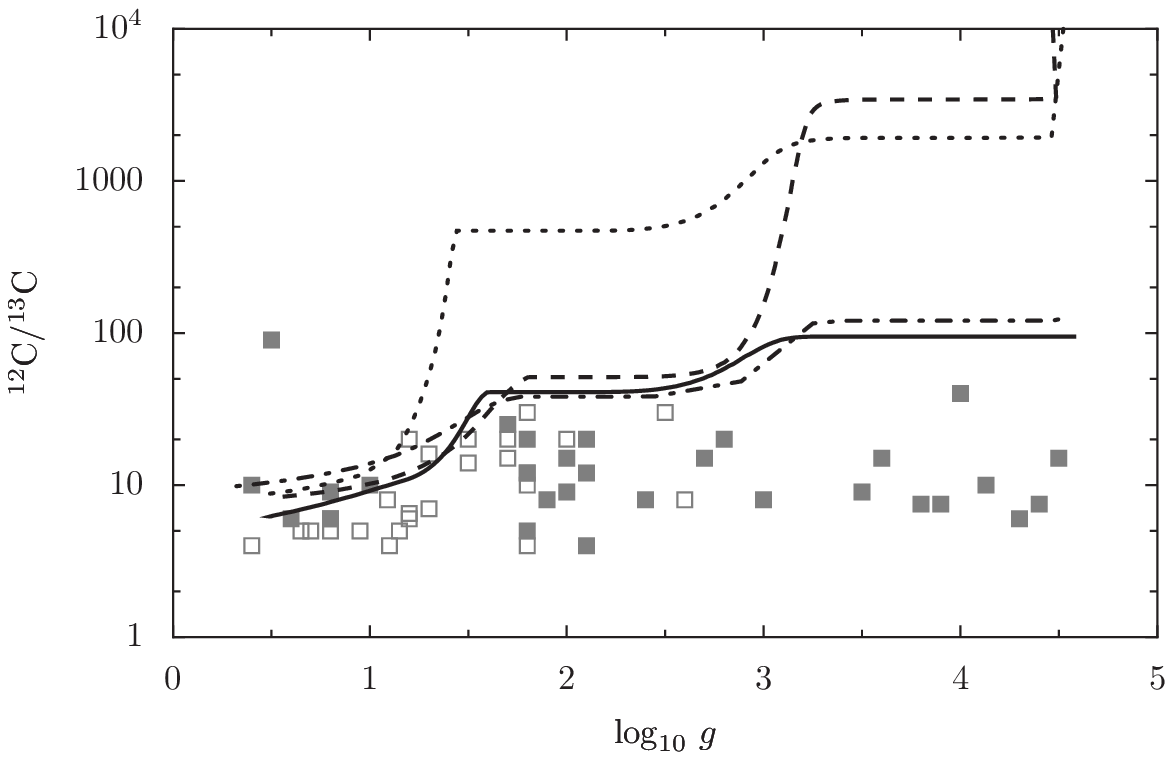}
\caption{Evolution of the \el{12}{C}/\el{13}{C} ratio as a function of $\log_{10} g$ in each of our models. The solid line indicates the carbon-normal model, the dotted line is the model which has accreted 0.01\ms\ of material and the dashed line is the model which has accreted 0.1\ms\ of material. The dot-dashed line is for a model accreting 0.1\ms\ of material that has a \el{12}{C}/\el{13}{C} ratio of around 100. The open squares represent C-normal stars, while the filled squares are for C-rich stars.}
\label{fig:12C13C}
\end{figure}

In Figure~\ref{fig:12C13C} we show the evolution of the \el{12}{C}/\el{13}{C} ratio as a function of surface gravity for our models. The models that have accreted material from an AGB companion initially have extremely high \el{12}{C}/\el{13}{C} ratios (of several thousand). At first dredge-up the ratio drops dramatically in the 0.1\ms\ accretion case because its accreted material has been mixed deeply into the star by thermohaline mixing and CN-cycled before being brought to the surface again. The 0.01\ms\ model shows a smaller drop in its \el{12}{C}/\el{13}{C} ratio because its material is not mixed as deeply (i.e. to those points in the star where CN-cycling is possible) and the material suffers only dilution during first dredge-up. The \el{12}{C}/\el{13}{C} ratio is substantially affected by the extra mixing on the upper part of the giant branch. In all cases, we find that this ratio is reduced to less than 10 in all our models by the time the top of the red giant branch is reached. This is unsuprising, given the quantity of nitrogen that is produced in these models. 

For the carbon-normal metal-poor stars, our model \el{12}{C}/\el{13}{C} ratios are in reasonable agreement with the observations of \citet{2006A&A...455..291S}, which are shown as crosses in Figure~\ref{fig:12C13C}. The post-first dredge-up abundances are towards the upper end of the observed ratios, but the decline owing to the extra mixing is well reproduced. For the CEMP stars, we note observations of the \el{12}{C}/\el{13}{C} ratio tend to be around 10 or less \citep{2006A&A...459..125S,2007AJ....133.1193B}. This is found across the whole range of surface gravity, i.e. in stars from the main sequence to the giant branch, and hence cannot be held as evidence in favour of giant branch mixing in CEMP stars. For those CEMP stars formed by mass transfer from an AGB companion, there would need to be an extra mixing mechanism in operation during the AGB in order to prevent the \el{12}{C}/\el{13}{C} ratios reaching the extremely high values predicted by our models. If a companion star accreted material of a much lower \el{12}{C}/\el{13}{C} ratio, it would evolve in a similar way to the models already presented. The dot-dashed line of Figure~\ref{fig:12C13C} shows what would happen to our model which accretes 0.1\ms\ of material if the \el{12}{C}/\el{13}{C} ratio of the ejecta were closer to the upper limit of the observations. The model suffers less of a drop in this ratio at first dredge-up because the envelope is closer to the equilibrium value expect from CN-cycling. The final \el{12}{C}/\el{13}{C} ratio reached at the tip of the giant branch is almost identical to that of the original model.

We note that our results are at odds with the work of \citet{2008ApJ...679.1541D}, who found it necessary to introduce a reduced efficiency of mixing to explain the observed abundance trends in CEMP stars. Their models produce a $\mu$ inversion in a layer of the star that is sufficiently cool that little CN-cycling takes place \citep[see the lower panel in fig. 6 of ][]{2008ApJ...684..626D}. As such, they find the mechanism is unable to affect the surface CN abundances, unless the mixing takes place to a greater depth. They postulate that if mixing is able to overshoot to around one pressure scale height below the $\mu$ inversion, then sufficient CN cycling can take place to affect the surface abundances. We note that of the previous implementations of thermohaline mixing by \citet{2008ApJ...677..581E} and \citet{2007A&A...467L..15C} did not need to include any such overshooting in order to reproduce observed abundance variations.

\begin{figure}
\includegraphics[width=\columnwidth]{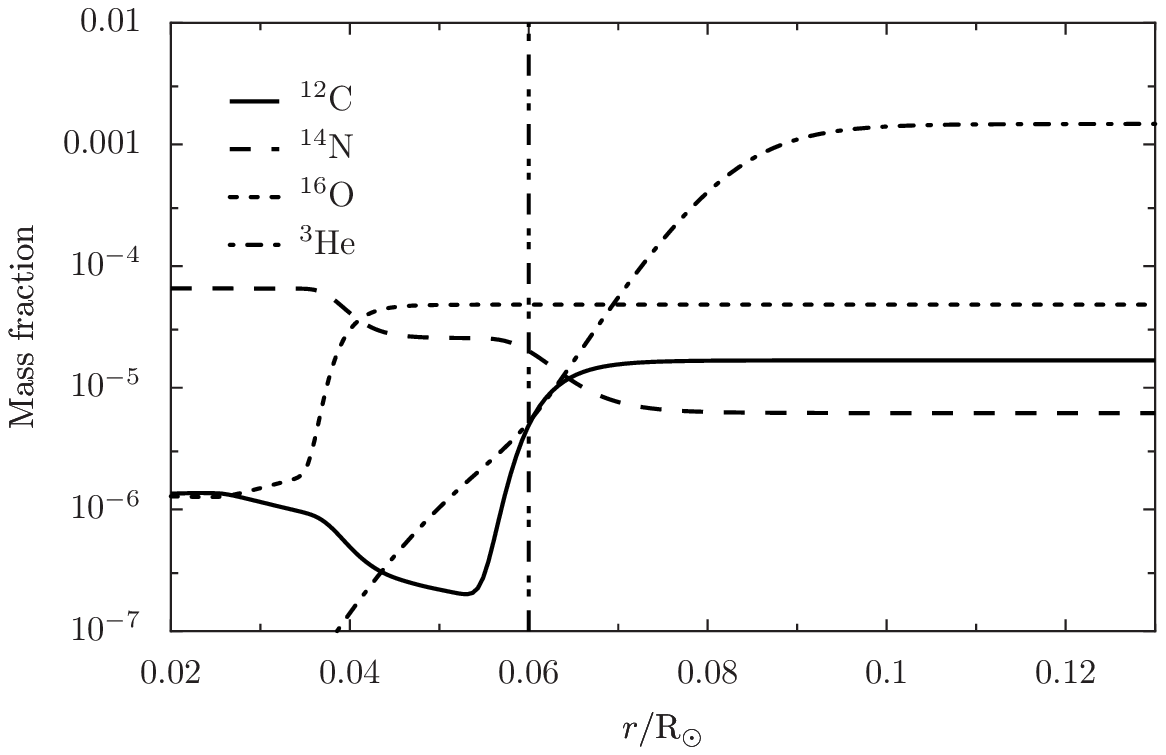}
\caption{Profiles of the mass fractions of the dominant CNO isotopes and \el{3}{He} as a function of radius for our 0.8\ms\ carbon-normal model, just prior to the onset of the extra mixing. The vertical line marks the radius at which the $\mu$-inversion develops.}
\label{fig:Profiles}
\end{figure}

In Figure~\ref{fig:Profiles} we plot the mass fraction profiles of the CNO elements and \el{3}{He} as a function of radius for our carbon-normal model. The profile is a very good match to that of Denissenkov \& Pinsonneault (cf. the lower panel of fig. 6 in their paper). We also confirm their finding that the radius at which the minimum $\mu$ is found does not vary much with time \citep{2008ApJ...679.1541D}. However, we find that our $\mu$ inversion develops at $r/\mathrm{R_\odot}=0.06$, which is slightly -- but significantly -- lower than their value of $r/\mathrm{R_\odot}\approx0.0675$. Crucially, this means our mixing is able to proceed to a depth where significant CN cycling occurs. 
We therefore do not need to invoke any overshooting beyond the point of minimum $\mu$ as Denissenkov \& Pinsonneault did. The reason for this variation in the location of the $\mu$ inversion is unclear. A detailed model comparison would be desirable, but insufficient information exists at present to make this possible. 

While $\delta\mu$-mixing appears to be able to describe the trends observed in both carbon-enhanced and carbon-normal metal-poor stars, we must add one note of caution. It has been pointed out that rotation could suppress this mixing mechanism \citep{2008ApJ...684..626D}.  While our code does not include rotational physics, it is not clear that simply adding the diffusion coefficients for both processes would give physical meaningful results. 3D hydrodynamic calculations of rotating fluid layers that are unstable to thermohaline mixing are clearly required.

\section{Conclusion}
We have investigated abundance changes in metal-poor giants in the context of the $\delta\mu$-mixing. We find that this mechanism accounts for the trends observed, namely that carbon-normal stars undergo significant depletion of carbon on the upper part of the red giant branch while carbon-rich stars do not. The reduced carbon depletion in CEMPs is a natural consequence of the proposed mixing mechanism and the efficiency of the mixing does not have to be reduced in an {\it ad hoc} way. For the carbon-normal model, the [N/Fe] abundance remains within the observed spread in metal-poor stars and the \el{12}{C}/\el{13}{C} ratio is reasonably well reproduced. The two models in which carbon-rich material has been accreted from a companion show substantial increases in nitrogen but their [N/Fe] values remain within the observed range. Their \el{12}{C}/\el{13}{C} ratios are brought down to around the observed value by the action of the $\delta\mu$ mixing.

The above results are presented with the caveat that rotation has not been taken into account in these simulations, and this may be able to supress the mixing. The interaction between thermohaline mixing and rotation (or {\it any} additional mixing mechanism) must be taken into account in future work and hydrodynamical simulations of such interactions are clearly warranted.

\section{Acknowledgements}
RJS and RPC are funded by the Australian Research Council's Discovery Projects scheme under grants DP0879472 and DP0663447 respectively. GCA is supported by a Faculty of Science Dean's Postgraduate Research Scholarship from Monash University.

\bibliography{../../../../masterbibliography}

\label{lastpage}
\end{document}